# Second harmonic generation induced by gate voltage oscillation in few layer MnBi$_2$Te$_4$

Liangcai Xu[1,*], Zichen Lian[1,*], Yongchao Wang[1,*], Xinlei Hao[1], Shuai Yang[2], Yongqian Wang[2], Chang Liu[2,3], Yang Feng[4], Yayu Wang[1,5,6], Jinsong Zhang[1,5,6,†]

[1]*State Key Laboratory of Low Dimensional Quantum Physics, Department of Physics, Tsinghua University, Beijing 100084, China*

[2]*Beijing Key Laboratory of Opto-electronic Functional Materials & Micro-Nano Devices, Department of Physics, Renmin University of China, 100872, Beijing, China*

[3]*Key Laboratory of Quantum State Construction and Manipulation (Ministry of Education), Renmin University of China, Beijing, 100872, China*

[4]*Beijing Academy of Quantum Information Sciences, Beijing 100193, China.*

[5]*Frontier Science Center for Quantum Information, Beijing 100084, China*

[6]*Hefei National Laboratory, Hefei, 230088, China*

* These authors contributed equally to this work.

† Emails: jinsongzhang@mail.tsinghua.edu.cn

**Abstract**

Nonlinear charge transport, such as nonreciprocal longitudinal resistance and nonlinear Hall effect, has attracted considerable interest in probing the symmetries and topological properties of new materials. Recent research has revealed significant nonreciprocal longitudinal resistance and nonlinear Hall effect in $MnBi_2Te_4$, an intrinsic magnetic topological insulator, induced by the quantum metric dipole. However, the inconsistent response with charge density and conflicting $C_{3z}$ symmetry requirement necessitate a thorough understanding of factors affecting the nonlinear transport measurement. This study uncovers an experimental factor leading to significant nonlinear transport signals in $MnBi_2Te_4$, attributed to gate voltage oscillation from the application of large alternating current. Additionally, a methodology is proposed to suppress this effect by individually grounding the voltage electrodes during second-harmonic measurements. The investigation underscores the critical importance of assessing gate voltage oscillation's impact before determining the intrinsic nature of nonlinear transport in 2D material devices with an electrically connected operative gate electrode.

**Introduction**

The observation of nonlinear Hall effect (NHE) in systems preserving time-reversal symmetry offers valuable insights into the underlying structure of the Berry curvature[1–5], distinct from the integral of Berry curvature revealed by the anomalous Hall effect (AHE)[6,7]. Furthermore, recent theoretical studies have revealed that the quantum metric can also induce a NHE[8,9]. This discovery is of great importance as it establishes a profound connection between quantum metric and NHE, underscoring the role of quantum geometry in condensed matter physics[10]. Specifically, quantum metric and Berry curvature are recognized as the real and imaginary parts of quantum geometry, respectively. While Berry curvature has been extensively studied and associated with various physics effects, such as the anomalous Hall effect[6,7], the quantum metric has received comparatively less attention because of its weak correlation with measurable physical quantities. In contrast to NHE, which is primarily linked to the topological properties of the Bloch wave of electrons, nonreciprocal longitudinal resistance (NLR) is typically attributed to the breaking of inversion symmetry[11], providing complementary insights into the interplay between topology and symmetry in quantum materials. Experimental observations of NLR have been reported in various systems, such as Bi helix[12], polar semiconductor BiTeBr[11], noncentrosymmetric superconductors[13–15], trigonal tellurium[16], Mott insulator $Ca_2RuO_4$[17], and quantum anomalous Hall effect system[18]. In addition to their significance in fundamental physics, NHE and NLR hold great potential for applications in the design of novel quantum materials and devices[19], such as high-frequency rectification[20] and terahertz detection[21].

In the intrinsic antiferromagnetic (AFM) topological insulator $MnBi_2Te_4$, recent experiments have demonstrated the existence of NHE and NLR, which are attributed to the quantum metric dipole owing to the unique *PT* symmetry (combine of parity and time-reversal symmetries)[22,23]. In this work, we report a previously overlooked measurement issue in 2D $MnBi_2Te_4$ devices, which can induce significant nonlinear charge transport signals, arising from the gate voltage oscillation caused by the application of a large alternating current (AC). Moreover, this issue has a general impact on all the NHE and NLR measurements of 2D material devices with operative gate electrode. Our study not only highlights the necessity of evaluating the contribution of gate voltage oscillation, but also provides a new paradigm in the nonlinear transport measurement of 2D material devices.

**Results**

**Nonlinear charge transport in $MnBi_2Te_4$**

$MnBi_2Te_4$ is an intriguing material known as an intrinsic AFM topological insulator[24–27]. It exhibits quantum anomalous Hall effect in the odd-number septuple-layer (SL) samples[28], characterized by dissipationless edge states with quantized Hall conductance, while it displays an axion insulator state in the even-SL sample, which is characterized by the presence of helical edge current with zero Hall plateau in low field region[29,30]. Our six-SL $MnBi_2Te_4$ flakes is exfoliated onto 285-nm $SiO_2$/Si substrates by using the Scotch tape method as reported previously[29]. Figure 1a illustrates the schematic diagram of our measurement setup. The AC is applied by Keithley 6221 current source with the frequency typically set at 13 Hz. Longitudinal voltage and transverse Hall voltage are simultaneously measured by two lock-in amplifiers connected to contacts 1-3 and contacts 1-2, respectively. The carrier density of the sample can be controlled by applying the gate voltage ($V_G$) using a Keithley 2400 source meter. In Fig. 1b, we present the current-voltage (IV) curves for both the longitudinal and transverse directions. The IV curve for the longitudinal direction shows sublinear behavior, which exhibits slight deviations from the previously reported results[22,23]. We attribute this discrepancy to the relatively high resistance in our sample, which can lead to voltage-induced scattering effects between the edges of the sample[31]. We also measured the IV curve by the direct current (DC) method, which is consistent with the AC method (see Supplementary Figure 2). Meanwhile, the transverse voltage $V_{yx}^{\omega}$ remains very small compared to $V_{xx}^{\omega}$, and the variation of $V_{yx}^{\omega}$ with $V_G$ will be discussed in Fig. 3.

As depicted in Fig. 1c, the observed loop behavior of the second-harmonic (2ω) Hall voltage ($V_{yx}^{2\omega}$) around zero magnetic field is believed to be closely related to two different AFM states in even-layer MnBi$_2$Te$_4$[22,23,32]. To be more specific, in the AFM I state, characterized by sweeping the magnetic field from -4 T to 0 T, the value of $V_{yx}^{2\omega}$ is positive. While in the AFM II state, where the magnetic field is swept from 4 T to 0 T, the value of $V_{yx}^{2\omega}$ is found to be negative (the raw data can be found in Supplementary Figure 5). Such loop, related to two different AFM states, is also observed in the first harmonic (1ω) Hall signal by tuning the electric field using a dual-gate device. This phenomenon is attributed to the opposite Berry curvature in adjacent layers[32].

As shown in Fig. 1d, the $V_{xx}^{2\omega}$ deviates from the quadratic behavior like that observed in other systems[11,12]. This departure can be attributed to the sublinear behavior of the IV curve within the high resistive region as discussed in supplementary materials (see Supplementary Note 4 and Supplementary Figure 3). In Fig. 1e, the nonlinear Hall voltage exhibits a quadratic scaling with the injected current $I^\omega$, consistent with previous reports[1,2,22,23]. To evaluate the strength of the NHE, we utilize the second harmonic of anomalous Hall voltage ($\Delta V_{yx}^{2\omega}$), which is calculated from the signals at AFM II and AFM I states: $\Delta V_{yx}^{2\omega} = \left(V_{yx}^{2\omega}(\text{AFM II}) - V_{yx}^{2\omega}(\text{AFM I})\right)/2$. The magnitude of $\Delta V_{yx}^{2\omega}$ reaches -0.25 mV at $I$ = 10 μA, which is comparable to the value in previous results[22,23]. Furthermore, it is evident in Fig. 1f that the signal gradually decreases as the temperature approaches the Neel temperature. Above the Neel temperature, the signal disappears within the limit of measurement.

During the experiments, we stumble upon an intriguing question: the nonlinear signals display a dramatic change when we alter the grounding contact. The measurement configuration employed in this study is illustrated in Fig. 2a. Here we introduce two switches, $S_S$ and $S_D$, to independently control the grounding. Figure 2b presents the measurement results of the NLR with different grounding contacts. When $S_S$ is switched on, the $V_{13}^{2\omega}$ value exhibits a dip followed by a peak as the $V_G$ increases. However, when $S_D$ is switched on ($S_S$ is disconnected), the $V_{13}^{2\omega}$ initially increases before transitioning to a negative value, which is nearly opposite to the results with the grounded source electrode. Meanwhile, the first harmonic signal of longitudinal voltage $V_{13}^\omega$ shows no difference between these two grounding conditions (see Supplementary Figure 6). This strongly indicates that the NLR arises from a measurement factor other than the intrinsic effect, such as the quantum metric dipole.

It is worth noting that the only alteration throughout the experimental procedure is the toggling of different switches. To clarify, the red clamp (positive electrode) of the current source (Keithley 6221) is consistently connected to the drain terminal, while the black clamp (negative electrode) is connected to the source terminal. The positive and negative input terminals of the lock-in amplifier are consistently connected to contact 1 and contact 3, respectively. As the output of the current source is isolated from the ground, we can ground the positive electrode. This unexpected behavior highlights the strong relevance between the grounding conditions and the resultant NLR response, which also suggests the possibility that a significant portion of the observed nonlinear signal is due to grounding artifacts rather than intrinsic material properties.

**Nonlinear charge transport induced by gate voltage oscillation**

In the following section, we will delve into the nonlinear signal using a simplified model. Notably, the simulation results obtained from this model exhibit consistent alignment with the experimental measurements.

At first, we start with discussing the situation of nonreciprocal longitudinal resistance. To ensure simplicity and clarity, we utilize positive and negative DC instead of AC. The equivalent of AC and DC is discussed in supplementary materials (see Supplementary Note 1). To illustrate the issue without losing generality, we consider the case when $S_S$ is switched on, thus the voltage potential of the source contact $V_S$ is set to zero.

When a positive current is applied (indicated by the red arrow in Fig. 2a), the voltage distribution on the sample is depicted by the red line in Fig. 2d. In this case, the voltage at contact 3 ($V_3$) is larger than the source electrode ($V_S = 0$ V). The voltage difference between electrodes 1 and 3 can reach about 0.2 V with 10-μA DC, as depicted in the IV curve of Fig. 1b. Furthermore, the effective gate voltage ($V_G^{eff}$) on the sample (the red line in Fig. 2e), which is defined as the difference between the applied $V_G$ and the voltage of the sample at different positions ($V_{sample}$), is smaller than the $V_G$ applied by Keithley 2400. In the general transport measurements, the difference between $V_G^{eff}$ and $V_G$ can be neglected. However, in the case of 6-SL MnBi$_2$Te$_4$, this discrepancy can be attributed to the presence of large longitudinal resistance, as illustrated by the IV curve in Fig. 1b and resistance peak in the inset of Fig. 2c, combining with the large current during second harmonic measurements.

However, when a negative current is applied (indicated by the green arrow in Fig. 2a), the voltage potential on the sample becomes negative (green line in Fig. 2d). Hence, the $V_G^{eff}$ on the sample is larger than the applied $V_G$ (green line in Fig. 2e). Consequently, the actual $V_G$ experienced by the sample is always oscillating when applying a large AC through the drain and source electrodes.

So far, we have demonstrated that the positive (negative) current can decrease

(increase) the $V_G^{eff}$. This behavior is clearly depicted in Fig. 2c, where the dash lines correspond to the simulated longitudinal voltage of DC measurements under the consideration of the shift of $V_G^{eff}$ as described in Fig. 2e. We find that the longitudinal resistance behaves as a right (left) shift for positive (negative) current compared to that measured by an AC of 10 µA. Besides that, the resistance is also sensitive to applied $V_G$ as displayed by the peak structure in the inset of Fig. 2c. These two typical characteristics lead to a sizable difference in longitudinal resistance between positive and negative currents. For example, the measured voltage $V_{13}$ has a difference of 6 mV between the 10-µA positive and negative currents, when the $V_G$ was set at 45 V, where the resistance changes dramatically. This difference reaches approximately one percent of the 1ω signal, which can serve as a distinctive characteristic of nonreciprocal charge transport.

The nonreciprocal feature becomes more evident in IV curves as depicted in Fig. 2f. For example, in the p-type region, the IV curve exhibits a reduced slope compared to the ideal linear IV curve with positive current. This arises from the decreased $V_G^{eff}$ and smaller resistance. In contrast, when the current is negative, the IV curve displays a heightened slope due to the increased $V_G^{eff}$ and larger resistance. This behavior can also be found in the n-type region, but the curvature is totally different due to the opposite sign of $dR_{xx}/dV_G$ compared to that in the p-type region. These non-antisymmetric IV curves are typical characteristics of nonreciprocal charge transport[33]. Moreover, when the grounding contact is switched from source to drain electrode, all the signs or trends discussed in Fig. 2c-2f would be reversed. By employing this simple model, we conducted simulations to reproduce the 2ω signal using the $R_{xx}$-$V_G$ data in Fig. 2c, which demonstrates a remarkable agreement with the experimental measurements (see Supplementary Figure 1).

In normal cases, the applied current is relatively small and the longitudinal voltage across the sample is regarded as a perturbation in the electrical transport measurements. However, in the case of nonreciprocal transport, the applied current is typically on the order of tens of micro-amperes, or even milliamps in some instances[11,33–35]. Such large currents can induce significant voltage differences between drain and source electrodes.

Furthermore, in 2D materials both the displacement field and charge carrier density can be dramatically tuned by the gate voltage[1,22,23], which can produce large resistance and considerable $dR_{xx}/dV_G$, consequently leading to pronounced signals in the nonreciprocal charge transport measurement. In further substantiating the universality of this phenomenon, we performed the NLR measurements in graphene devices to explore the impact of gate voltage oscillation. The outcomes are consistent with the deductions drawn from the MnBi$_2$Te$_4$ devices (see Supplementary Note 12).

Subsequently, we assess the influence of gate voltage oscillation on the NHE. As previously mentioned, even though the zero-field Hall voltage is relatively small compared to the longitudinal voltage, it still exhibits variations at different $V_G$. As shown in Fig. 3a, the voltage of the AHE at zero magnetic field, represented by $\Delta V_{yx}^{\omega} = \left(V_{yx}^{\omega}(\text{AFM II}) - V_{yx}^{\omega}(\text{AFM I})\right)/2$, displays a sign reversal behavior with the increase of $V_G$ and approaches to zero in the heavily n- and p-doped regions. The observed AHE in our sample is consistent with the layer Hall effect reported in MnBi$_2$Te$_4$, which is attributed to the opposite Berry curvature in adjacent layers[32]. However, for the second harmonic measurements, as depicted in Fig. 3b, the loop of nonlinear Hall voltage $V_{yx}^{2\omega}$ is completely inverted when switching the ground electrode between the drain and source contacts. This is in stark contrast to the negligible differences observed in the $1\omega$ anomalous Hall voltage $V_{yx}^{\omega}$. This feature is incomprehensible in the framework of quantum metric dipole[22,23], because such an intrinsic quantum effect should not rely on the grounding electrode. However, this phenomenon can be clearly explained by the gate voltage oscillation picture and the $2\omega$ signals are determined by both the gate voltage shift and the value of $d\Delta V_{yx}^{\omega}/dV_G$. When the grounding electrode is changed, the gate voltage shifts towards the opposite direction, resulting in a sign reversal of $2\omega$ Hall voltage. Moreover, Fig. 3c demonstrates that when one of the Hall voltage electrodes (labeled as 1 and 2) is grounded, the amplitude of $\Delta V_{yx}^{2\omega}$ is smaller compared to that with grounded drain or source electrode. This observation can be attributed to the reduced gate voltage oscillation in the sample region around Hall electrodes because the voltage of this region is closer to the zero potential of the ground when electrode 1 or 2 is grounded. However, it is noteworthy that the $1\omega$ data remains unaffected by different grounding conditions (see Supplementary Figure 6).

To suppress the influence of gate voltage oscillation, we take the average of $\Delta V_{yx}^{2\omega}$ obtained by grounding Hall electrodes 1 and 2 separately. The compensated result $\Delta V_{yx,comp}^{2\omega} = \left(\Delta V_{yx}^{2\omega}(\text{GND}_1) + \Delta V_{yx}^{2\omega}(\text{GND}_2)\right)/2$, is illustrated in Fig. 3d. This calculation is reasonable because the contribution of gate voltage oscillation is totally opposite when grounding electrodes 1 and 2. Furthermore, the current following through the Hall electrodes is completely negligible and it is not necessary to consider the difference in contact resistance of each Hall electrode. In the cases of grounding source or drain electrode, the contact resistance would affect the magnitude of gate voltage oscillation. However, it is worth noting that, even after the compensation, the second harmonic signals still exist, accounting for approximately one fifth of the original signal. It is plausible that these signals could be attributed to the crossed nonlinear dynamical Hall effect, which is driven by the combined influence of in-plane and time-varying out-of-plane alternating current fields[36]. Nevertheless, additional experiments are necessary to validate this hypothesis or the quantum metric effect.

**Discussion**

To further illustrate the above picture, we propose a fresh perspective to shed light on this problem by intentionally introducing gate voltage oscillation, which has the same frequency and is in phase with AC following through source and drain electrodes. As illustrated in Fig. 4a, the gate voltage applied on the sample includes two parts. The DC part is applied by Keithley 2400, and the AC part is generated by a voltage transformer with a primary-to-secondary voltage ratio of 4:1[37]. As illustrated in Fig. 4b, the NLR signal is negligible when applying a small current $I = 0.2$ μA. However, when the primary voltage of the transformer is set to 5 V, which induces a gate voltage oscillation of 1.25 V, a substantial NLR signal is detected in both p-type and n-type regions.

After successfully demonstrating the generation of nonreciprocal transport by the intentional gate voltage oscillation, we conduct measurements using a larger AC. This allows us to superpose the magnitude of gate voltage oscillation from both the large AC and the transformer. The primary voltage of the transformer is varied from 0 V to 5 V, as shown in Fig. 4c. Notably, all the curves intersected at a single point (-0.5 mV) near the charge neutrality point (CNP), where the nonreciprocal transport signals are expected to be zero due to the zero slope of $V_{xx}^\omega$ around the resistance peak, i.e., $dV_{xx}^\omega/dV_G = 0$ at CNP. Further experiments are required to investigate the origin of this deviation. Additionally, we also observed that the application of intentional gate voltage oscillations can reverse the second harmonic signals of the nonlinear Hall effect, as depicted in Fig. 4d. Specifically, when the primary voltage is set to 3 V, the $2\omega$ Hall voltage exhibits no loop, indicating successful suppression of gate voltage oscillation arising from both the AC and the transformer.

In conclusion, we observed strong nonreciprocal longitudinal resistance and nonlinear Hall effect induced by gate voltage oscillation during the second harmonic measurements in MnBi$_2$Te$_4$ and graphene. In all the previous nonlinear measurements of 2D material devices, only the source electrode (negative electrode of current source) is grounded and the acquired data always contain the contribution of gate voltage oscillation, which may invalidate all the analysis and discussions about the intrinsic mechanisms. Our observations indicate that this artifact becomes more dominate in the following two conditions. Firstly, the voltage drop across the sample should be large enough to shift the applied gate voltage, which can be realized in the samples with large resistance and large applied current Secondly, the signal (including longitudinal resistance and Hall resistance) should be sensitive to the gate voltage, which allows detectable resistance differences for positive and negative currents. This phenomenon is particularly pronounced in hBN-gated devices due to the large geometry capacitance of gate electrode. Moreover, we propose a method to suppress the effect of gate voltage oscillation by individually grounding the electrodes for voltage measurements. Our study emphasizes the crucial importance of evaluating the contribution of gate voltage oscillation before drawing definitive conclusions regarding the intrinsic nature of nonlinear signals in 2D material devices that can be effectively controlled by an in-situ gate electrode.

**Methods**

**Crystal growth** The MnBi$_2$Te$_4$ single crystal was grown by the chemical vapor transport (CVT) method[38–40]. Mn (99.95%, Alfa Aesar), Te (99.999%, Alfa Aesar), and Bi$_2$Te$_3$ (99.999% Alfa Aesar) lumps were meticulously mixed in a 2:2:1 stoichiometric ratio, ground into powder, and placed in a quartz ampoule along with iodine (I$_2$, 99.99%, 3A) as transport agent in a proportion of 0.3 times the stoichiometric ratio. The sealed quartz ampoule was held at 900 °C over 6 hours ensuring the raw materials were well mixed in a box furnace. Following a controlled cooling process, the sealed quartz ampoule was placed in a tube furnace with a controllable gradient temperature for the CVT-driven crystal growth process. The temperature of the source end and growth end were held at 597 °C and 588 °C respectively for 30 days. Then the quartz ampoule was quenched in water to prevent the side phases. Millimeter-sized MnBi$_2$Te$_4$ single crystals can be found at the growth end. The crystal structure and magnetic order of crystals were confirmed through X-ray diffraction (XRD) and magnetic measurements.

**Device fabrication** MnBi$_2$Te$_4$ flakes were exfoliated onto an SiO$_2$/Si wafer, obtained through thermal oxidation of silicon, with a SiO$_2$ thickness of 285 nm, by using the Scotch tape method in an argon-filled glove box with O$_2$ and H$_2$O levels lower than 0.1 ppm. Before exfoliation, all SiO$_2$/Si substrates were pre-cleaned by air plasma for 5 minutes at ~ 125 Pa pressure. For the transport devices, thick flakes around the target sample were scratched off by using a sharp needle in the glove box. A layer of 270 nm PMMA was spin-coated before electron beam lithography (EBL) and heated at 60 °C for 5 minutes. After the EBL, Cr/Au electrodes (3nm/20-50nm) were deposited by a thermal evaporator connected with an argon-filled glove box. Before the fabrication and sample transfer process, the devices were always spin-coated with a PMMA layer to avoid exposure to air. The six-SL MnBi$_2$Te$_4$ has the thickness of 8.2nm[29], the width and length of the sample has been summarized in Supplementary Table 1.

**Transport measurement** Four probe transport measurements were carried out in a cryostat with the lowest temperature of 1.4 K and out-of-plane magnetic field up to 8 T. The first and second harmonic longitudinal and Hall signals were acquired simultaneously via lock-in amplifiers (SR830) with an AC (0.1~10 μA, ~13 Hz) generated by a Keithley 6221 current source. To correct for the geometrical misalignment, the longitudinal and Hall signals were symmetrized and anti-symmetrized with the magnetic field respectively. The back gate voltages were applied by a Keithley 2400 source meter.

**Data Availability:** All raw and derived data used to support the findings of this work are available from the authors on request.

**Acknowledgments:** This work is supported by the Basic Science Center Project of NSFC (grants No. 52388201), and the National Natural Science Foundation of China (grant No. 12350404, 12274252, 12274453). We thank the financial support from the Innovation Program for Quantum Science and Technology (Grant No. 2021ZD0302502). Liangcai Xu is supported by a project funded by China Postdoctoral Science Foundation (No. 2021M701947) and the Shuimu Tsinghua Scholar Program. Chang Liu is supported by Open Research Fund Program of the State Key Laboratory of Low-Dimensional Quantum Physics (Grant No. KF202204).

**Author contributions:** J. S. Z., Y. Y. W., Y. F., and C. L. supervised the research. L. C. X., Z. C. L., Y. C. W., X. L. H., S. Y., and Y. Q. W fabricated the devices and performed the transport measurements. Y. C. W. grew the MnBi$_2$Te$_4$ crystals. L. C. X. and J. S. Z. prepared the manuscript with comments from all authors.

**Competing interests**

The authors declare no competing interests

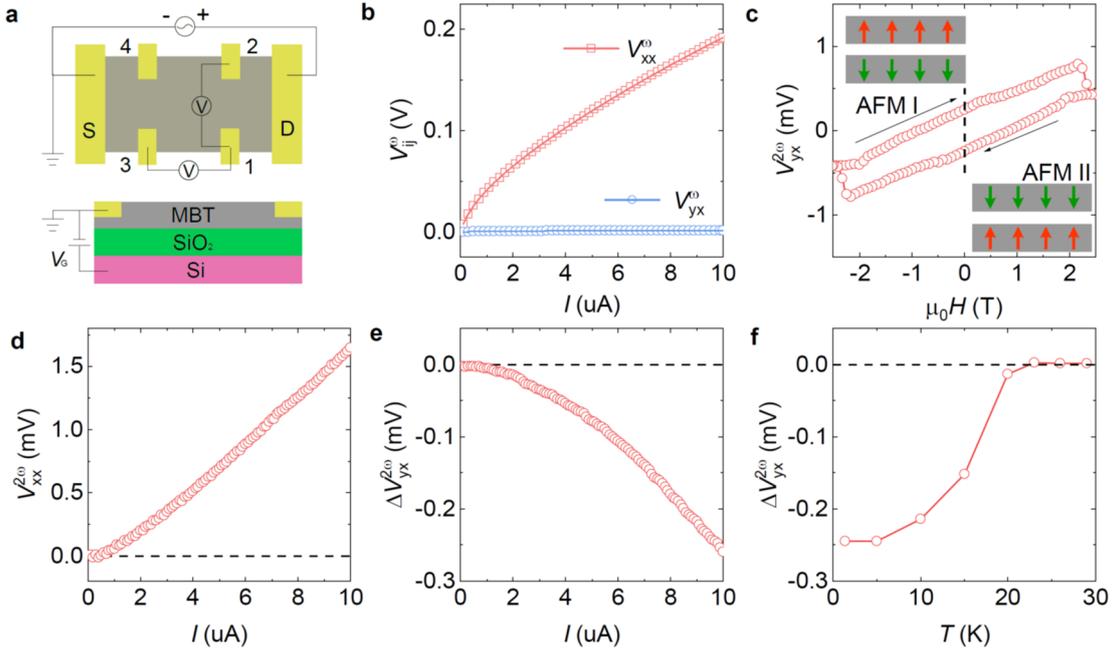

**Fig. 1 | Nonlinear charge transport in MnBi$_2$Te$_4$. a**, Schematic diagram of the device and measurement setup. The six-SL MnBi$_2$Te$_4$ flake was exfoliated on a 285-nm SiO$_2$/Si substrate. **b**, Current dependent first harmonic (1ω) longitudinal and transverse voltage. **c**, Second harmonic (2ω) transverse voltage while sweeping up/down magnetic field. The Inset carton indicates two different states (AFM I/II) at zero magnetic field, respectively. AFM I, sweeping the magnetic field from -4 T to 0 T; AFM II, sweeping the magnetic field from +4 T to 0 T. **d**, Current-dependent 2ω longitudinal voltage. **e**, The difference of 2ω transverse voltage ($\Delta V_{yx}^{2\omega}$) between AFM II and AFM I at different currents. **f**, The temperature evolution of $\Delta V_{yx}^{2\omega}$. All the data are acquired with $V_G$ = 75 V and $T$ = 1.4 K for (b-e).

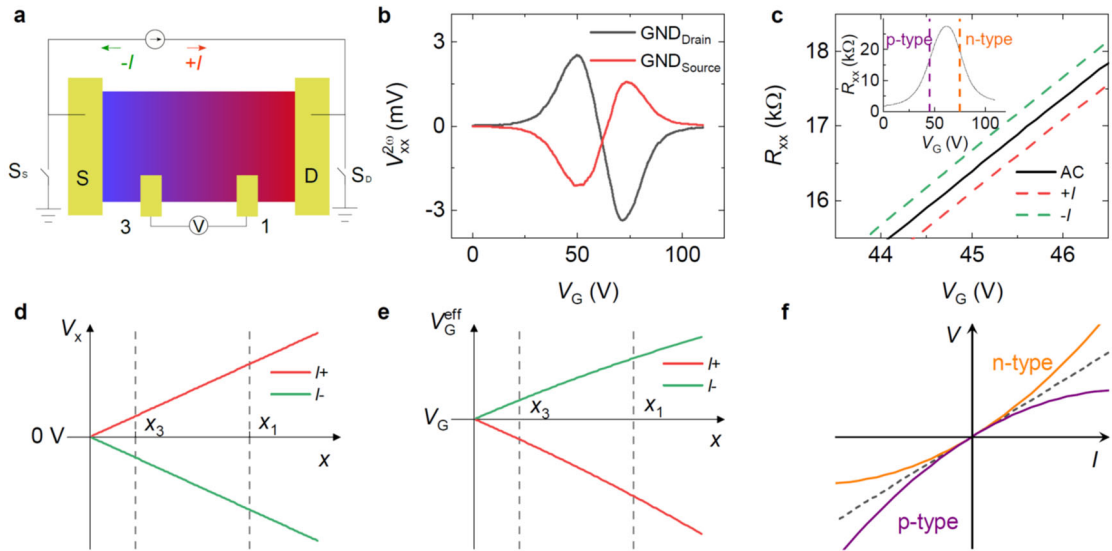

**Fig. 2 | Nonreciprocal longitudinal resistance induced by gate voltage oscillation. a**, Measurement configuration: the red clamp of the Keithley 6221 current source is consistently connected to the drain terminal, while the black clamp was connected to the source terminal. During the measurements, only one switch ($S_S$, $S_D$) is closed to control the grounding electrode (source or drain), while the other one remains open. The color illustrates the spatial distribution of effective gate voltage on the device. **b**, Second harmonic signals of NLR obtained by grounding drain or source terminal respectively. **c**, Gate-voltage dependence of longitudinal resistance (black solid line) around the p-type region measured by using AC $I = 10$ μA at $T = 1.4$ K. The inset shows the whole profile of resistance peak. The dashed red/green line represent the shift induced by applying positive/negative direct current. **d**, Longitudinal voltage distribution as a function of the position within the sample for positive current (red line) and negative current (green line). **e**, The distribution of effective gate voltage within the sample for different current directions. **f**, The IV curves of n-type and p-type regions due to the shift of gate voltage. The source electrode is grounded in (c-f).

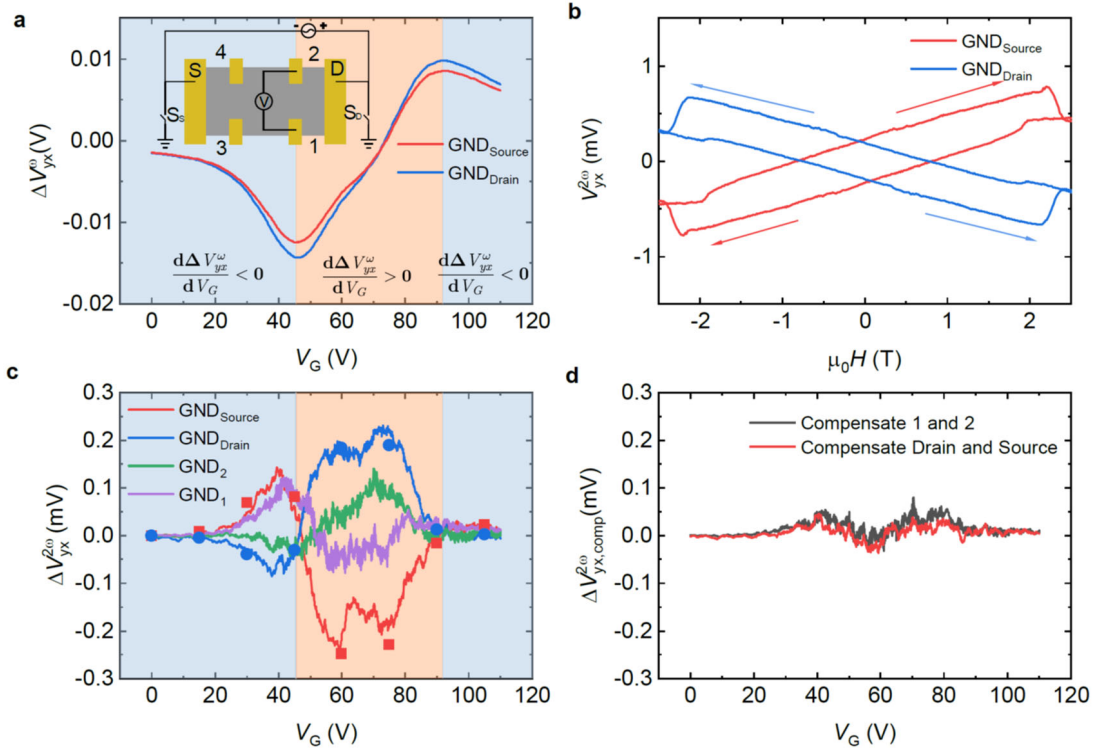

**Fig. 3 | Nonlinear Hall effect with different grounding contacts. a**, The gate-dependent anomalous Hall voltage $\Delta V_{yx}^{\omega}$ with grounding drain and source contact separately. The two colored regions represent the negative and positive slopes of $\Delta V_{yx}^{\omega}$ versus gate voltage, respectively. The measurement configuration is illustrated in the inset. **b**, Nonlinear Hall voltage $V_{yx}^{2\omega}$ obtained by sweeping the magnetic field with the same grounding conditions shown in a. **c**, The gate-dependent second harmonic Hall voltage $\Delta V_{yx}^{2\omega}$ with grounding drain, source, contact 1, and contact 2 separately. The red and blue solid dots correspond to the data obtained from the $\Delta V_{yx}^{2\omega}$ curves by sweeping the magnetic field, as shown in b. **d**, The compensated data of second harmonic Hall voltage $\Delta V_{yx,comp}^{2\omega}$ obtained by averaging the paired curves in c with grounding Hall electrodes (1 and 2) or source and drain electrodes separately. All the data are acquired with $T = 1.4$ K.

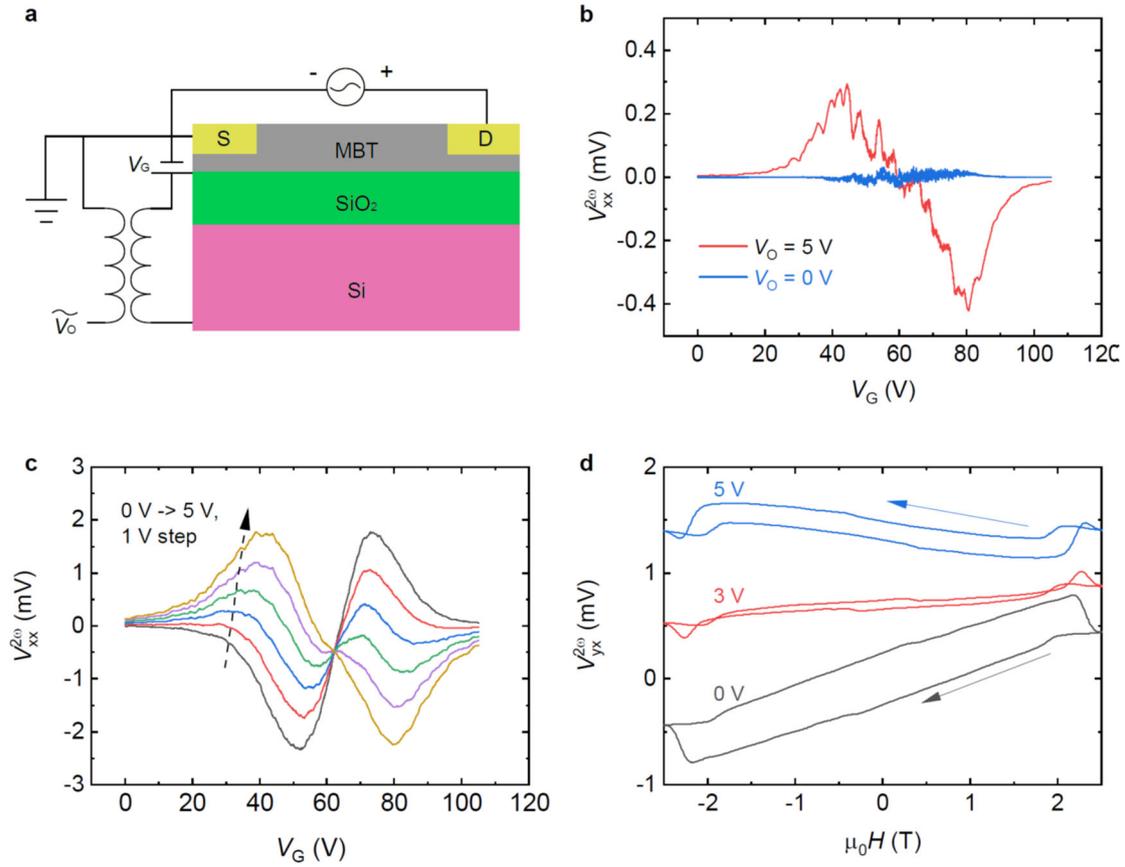

**Fig. 4 | Nonlinear charge transport with intentional gate voltage oscillation. a**, Measurement setup for the application of intentional gate voltage oscillation with a voltage transformer. The AC component of gate voltage has the same frequency and phase as the drain-source current. **b**, Gate-voltage-dependent NLR with applied AC $I$ = 0.2 μA. The primary voltage of the transformer is set to 0 V and 5 V respectively. **c**, Gate-dependent NLR with applied AC $I$ = 10 μA. The primary voltage is varied from 0 V to 5 V. **d**, The second-harmonic Hall voltage versus the magnetic fields. The loops are offset by 0.7 mV for clarity. Gate-dependent anomalous second harmonic Hall voltage with applied current $I$ = 10 μA, the primary voltage is set to 0, 3, and 5 V respectively.